\documentclass[%
 reprint,
showpacs,
 amsmath,amssymb,
 aps,
 prl,
]{revtex4-1}

\usepackage{graphicx}% Include figure files
\usepackage{dcolumn}% Align table columns on decimal point
\usepackage{bm}% bold math
\usepackage{amsthm, enumitem, caption, subcaption}
\usepackage{tikz}
\usetikzlibrary{arrows,automata}
\usetikzlibrary{decorations,decorations.pathmorphing}

\newcommand{\Z}{\mathbb{Z}}

\newcommand{\cE}{\mathcal{E}}

\newcommand{\cI}{\mathcal{I}}
\newcommand{\cO}{\mathcal{O}}

\renewcommand{\Pr}{\mathbb{P}}
\newcommand{\1}{\mathbf{1}}

\DeclareMathOperator\dist{dist}

\newcommand{\extB}{\partial_{\circ}}
\newcommand{\intextB}{\partial_{\ins \out}}
\newcommand{\ins}{\bullet}
\newcommand{\out}{\circ}
\newcommand{\bad}{\mathsf{bad}}

\newcommand{\overlap}{\mathsf{overlap}}

\newtheorem{thm}{Theorem}

\newcommand{\zero}{{\bf 0}}

\begin{document}

%\preprint{APS/123-QED}

\title{A condition for long-range order in discrete spin systems with application to the antiferromagnetic Potts model}% Force line breaks with \\

\author{Ron Peled}
\email{peledron@post.tau.ac.il}
\author{Yinon Spinka}%
 \email{yinonspi@post.tau.ac.il}
\affiliation{%
School of Mathematical Sciences, Tel Aviv University
}%

\date{\today}% It is always \today, today,

\begin{abstract}
We give a general condition for a discrete spin system with nearest-neighbor interactions on the $\Z^d$ lattice to exhibit long-range order. The condition is applicable to systems with residual entropy in which the long-range order is entropically driven. As a main example we consider the antiferromagnetic $q$-state Potts model and rigorously prove the existence of a broken sub-lattice symmetry phase at low temperature and high dimension -- a new result for $q\ge 4$. As further examples, we prove the existence of an ordered phase in a clock model with hard constraints and extend the known regime of the demixed phase in the lattice Widom-Rowlinson model.
\end{abstract}

\pacs{64.60.Cn, 05.50.+q, 75.10.Hk, 64.60.De}

\maketitle

{\it Introduction}.---The nature of the low-temperature phase in classical lattice spin systems, such as the ferromagnetic Ising or Potts models, is well-understood, with the system exhibiting a sea of aligned spins interrupted by small islands
of other spin values. Indeed, the classical Peierls argument compares the energetic cost of creating a domain wall between different spin values with the entropy of the wall's location and concludes that such walls are strongly disfavored at low temperatures.

The behavior at low temperature, indeed even at zero temperature, is less clear for systems with residual entropy. The antiferromagnetic (AF) $q$-state Potts model on the hypercubic lattice $\Z^d$ is a representative example. When $q$ is large compared with $d$ ($q>4d$ suffices~\cite{salas1997absence,Dobrushin1968TheDe}), the system is disordered at all temperatures, including zero temperature. In the opposite regime, when $q$ is fixed and $d$ large, say, Berker and Kadanoff~\cite{berker1980ground} initially suggested that a phase with algebraically decaying correlations may occur at low temperature. This was followed by numerical simulations and an $\varepsilon$-expansion argument of Banavar, Grest and Jasnow~\cite{banavar1980ordering} who predicted a \emph{broken-sublattice symmetry} (BSS) phase at low temperatures for the $3$ and $4$-state models in three dimensions. In the BSS phase, the densities of the different spin values differ between the two sublattices, with $\lfloor \frac{q}{2}\rfloor$ states dominant on one sublattice and $\lceil \frac{q}{2}\rceil$ dominant on the other (though there are excitations even at zero temperature). Koteck\'y~\cite{kotecky1985long} further argued for the existence of the BSS phase by relating the model to the ferromagnetic case. The BSS phase is an example of an entropically-driven long-range order, in which spins tend to take values in a restricted set of possibilities on one sublattice in order to allow more possibilities for the spins on the other sublattice.

It is expected that for each lattice $\mathcal{L}$ there is a value $q_{\text{c}}(\mathcal{L})$ such that the AF Potts model orders at low temperatures for $q<q_{\text{c}}(\mathcal{L})$ and is disordered at all temperatures for $q>q_{\text{c}}(\mathcal{L})$. However, it is difficult to rigorously establish the existence of entropically-driven long-range order, as the standard Peierls argument is obstructed by the fact that domain walls have no energetic contribution but rather serve only to reduce the entropy of the configuration around them. Thus, a rigorous proof of existence of the BSS phase on the $\Z^d$ lattice is available only for $q=3$ and $d$ large~\cite{peled2010high,galvin2012phase,feldheim2015long} (so that $q_{\text{c}}(\Z^d)\ge 3$ for $d$ large). Irregularities in the lattice $\mathcal{L}$, i.e., having different sublattice densities, promote ordering. The $q=3$ ($q=4$) model is disordered at all temperatures on the honeycomb lattice \cite{shrock1997ground, salas19983} and has a zero-temperature critical point on the square~\cite{lieb1967residual, salas1998three} (triangular~\cite{nienhuis1982exact, baxter1986q, moore2000height}) lattice, but orders at low temperatures on the so-called diced~\cite{feldmann1997mapping, kotecky2008phase, kotecky2014entropy} (union-jack~\cite{chen2011partial}) lattice.
More extreme irregularities can increase $q_{\text{c}}$ to arbitrary large values on planar lattices~\cite{huang2013two}. However, irregularities also modify the nature of the resulting phase, leading to long-range order in which a single spin value appears on most of the lower-density sublattice \cite{kotecky2014entropy}, or to partially ordered states \cite{qin2014partial}.

In this letter, we study the AF Potts model on the hypercubic lattice and give a rigorous proof of existence of a BSS phase for any number of states $q$, when the dimension is taken sufficiently high. More precisely, we prove that $q_{\text{c}}(\Z^d) \ge c\,d^{1/20}\log^{-2/5}d$ for a universal constant $c>0$. Together with the bound $q_c(\Z^d)\le 4d$~\cite{salas1997absence} this shows power-law growth of $q_c$ in high dimensions. Our techniques apply more generally to discrete spin systems with nearest-neighbor isotropic interaction possessing certain symmetries and our main result is an explicit condition on such systems ensuring the existence of long-range order.

\smallskip
{\it Model}.---We consider a general spin system on $\Z^d$ with spins taking values in a set $S$ of size $q$. The temperature-dependent probability of a configuration $f\colon\Lambda\to S$ in a finite domain $\Lambda\subset\Z^d$ can be expressed as
\begin{equation*}
  \Pr_{\Lambda}(f) = \frac{1}{Z_{\Lambda}}\prod_{v \in \Lambda} \lambda_{f(v)} \prod_{u,v \in \Lambda\text{ n.n.}} \lambda_{f(u),f(v)} ,
\end{equation*}
where $(\lambda_i)$ are positive activities, $(\lambda_{i,j})$ are non-negative symmetric nearest-neighbor interactions and $Z_\Lambda$ is a normalizing partition function. Boundary conditions may be further imposed. This abstract model is very general, allowing any discrete spin system with external magnetic fields, temperature parameter and nearest-neighbor symmetric interactions. Models with hard-core interaction, as arise at zero temperature, are included, by setting some of the $\lambda_{i,j}$ to zero.
The $q$-state AF Potts model at inverse temperature $\beta>0$ is obtained when
\[ S = \{1,\dots,q\},\quad\lambda_i=1,\quad\lambda_{i,j}=\1_{\{i\neq j\}}+e^{-\beta}\1_{\{i=j\}} ,\]
where $\1_{E}$ equals $1$ when $E$ holds and equals $0$ otherwise.

As discussed, our main result is a condition for such a system to exhibit long-range order. We start by explaining the type of ordering which may arise. Set
\begin{equation*}
  \lambda_{\text{max}} := \max_{i,j} \lambda_{i,j}.
\end{equation*}
A \emph{pattern} is a pair $(A,B)$ of subsets of $S$ such that $\lambda_{a,b}=\lambda_{\text{max}}$ for all $a \in A$ and $b \in B$.
The weight of a pattern $(A,B)$ is $\lambda_A \lambda_B$, where $\lambda_U := \sum_{u \in U} \lambda_u$ for $U \subset S$. Set
\begin{equation*}
   w_{\text{max}}:=\max_{(A,B)\text{ pattern}} \lambda_A \lambda_B.
\end{equation*}
A pattern is \emph{dominant} if it has weight $w_{\text{max}}$. The system will exhibit an ordering of the following kind: in a typical realization, a dominant pattern $(A,B)$ is chosen and then most spins on one sublattice take values in $A$ and most spins on the other sublattice take values in $B$. As an example, the BSS phases occuring for the AF Potts model are of this type with $(A,B)$ forming a partition of the $q$ states into $\lfloor \frac{q}{2}\rfloor$ and $\lceil \frac{q}{2}\rceil$.

\smallskip
{\it Conditions.}---We postulate that the internal symmetry group of the system acts transitively on the dominant patterns and a quantitative inequality involving the activities, interaction and dimension ensuring that excitations of the dominant patterns are sufficiently suppressed.

The system satisfies (SYM) if any two dominant patterns $(A,B)$ and $(A',B')$ are equivalent in the sense that there exists a bijection $\varphi \colon S \to S$ such that
\begin{equation*}
  \text{$\{\varphi(A),\varphi(B)\} = \{A',B'\}$, $\lambda_{\varphi(i)}=\lambda_i$, $\lambda_{\varphi(i),\varphi(j)}=\lambda_{i,j}$}
\end{equation*}
for all $i,j \in S$. We emphasize that if $(A,B)$ is a dominant pattern with $A\neq B$, then $(A,B)$ and $(B,A)$ are two equivalent, albeit distinct, dominant patterns.

The system satisfies (GAP), with constants $C_0,c_0$, if
\begin{equation*}
  \log \left[ \frac{1}{1 - (1-\rho_{\text{p}})(1-\rho_{\text{i}}^{c_0/q})} \right] \ge \frac{C_0q^3\log^2 d + \log \frac{1}{\rho_{\text{a}}}}{d^{1/4}},
\end{equation*}
where $\rho_{\text{a}}, \rho_{\text{i}}$ and $\rho_{\text{p}}:=\max\{\rho_{\text{p}}^1, \rho_{\text{p}}^2\}$ are the activity, interaction and pattern ratios defined as follows:
\begin{align*}
  &\rho_{\text{a}} := \frac{\min_i \lambda_i}{\max_i \lambda_i}, &&\rho_{\text{p}}^1 := \max_{\substack{(A,B)\text{ non-dom}\\\text{max pattern}}} \tfrac{\lambda_A \lambda_B}{w_{\text{max}}},\\
  &\rho_{\text{i}} := \max_{\substack{i,j\\\lambda_{i,j}<\lambda_{\text{max}}}} \tfrac{\lambda_{i,j}}{\lambda_{\text{max}}}, &&\rho_{\text{p}}^2 := \max_{\substack{(A,B)\text{ dom pattern}\\(A',B')\text{ max pattern}\\A' \subsetneq A}} \tfrac{\lambda_{A'}}{\lambda_A},
\end{align*}
where a pattern $(A,B)$ is maximal if no other pattern $(A',B')$ satisfies $A\subset A'$ and $B\subset B'$. It is worth noting that any $(\lambda_i)$ and $(\lambda_{i,j})$ satisfy (GAP) in sufficiently high dimensions.

\smallskip
{\it Results}.---A vertex of $\Z^d$ is \emph{even} or \emph{odd} according to the parity of the sum of its coordinates.
Given a configuration $f$, a vertex $v$ \emph{follows the $(A,B)$-pattern} if $v$ is even and $f(v) \in A$, or $v$ is odd and $f(v) \in B$. A set of vertices is an \emph{$(A,B)$-cluster} if it is a maximal connected set of vertices following the $(A,B)$-pattern.
\begin{thm}\label{thm:main}
There exist $C_0,c_0>0$ such that the following holds. Suppose $d\ge 2$ and the spin system satisfies (SYM) and (GAP). Then for any dominant pattern~$(A,B)$, the system exhibits an ordered phase characterized by having, almost surely, a unique infinite $(A,B)$-cluster and having no infinite $(A',B')$-cluster for any other dominant pattern $(A',B')$. These phases are extreme equilibrium states, invariant to automorphisms preserving the two sublattices, and the system has no other extreme, periodic, equilibrium states.
\end{thm}
Applied to the AF Potts model, Theorem~\ref{thm:main} shows that when $q \le c\,d^{1/20}\log^{-2/5}d$ and $\beta \ge Cq^6 d^{-1/4} \log^2 d$, there are $\binom{q}{q/2}$ or $2\binom{q}{\lfloor q/2 \rfloor}$ extreme equilibrium states according to whether $q$ is even or odd.

The proof of Theorem~\ref{thm:main} reveals that the global structure of the $\Z^d$ lattice is not essential to the result. The proof adapts to other lattices of coordination number at least $d$ and dimension at least $2$ which have some of the local features of $\Z^d$, such as the lattice $\Z^2\times\{0,1\}^{d-2}$; see~\cite{peledspinka17} for details.

\smallskip
{\it Examples}.---The lattice \emph{Widom-Rowlinson model} at activity $\lambda$ is given by
\[ S=\{-1,0,1\},\quad\lambda_i=\lambda^{|i|},\quad\lambda_{i,j}=\mathbf{1}_{\{ij \neq -1\}} .\]
The dominant patterns are $(\{0,1\},\{0,1\})$ and $(\{0,-1\},\{0,-1\})$,
and Theorem~\ref{thm:main} shows that when $\lambda \ge Cd^{-1/8} \log d$, there are two extremal periodic Gibbs states, characterized by an unequal density of $\pm 1$. This was previously known only when $\lambda \ge e^{Cd}$~\cite{lebowitz1971phase,burton1995new}.

\smallskip
The \emph{$\Z_q$-clock model with `hammock' potential of width $m$} is obtained when
\[ S=\Z_q,\quad \lambda_i=1,\quad \lambda_{i,j} = \mathbf{1}_{\{\dist_{\Z_q}(i,j) \le m \}} .\]
When $m <\frac{q}{4}$, the dominant patterns are $(i+A,i+A)$, $i\in S$, $A:=\{ 0,1,\dots,m \}$. Theorem~\ref{thm:main} then shows that when $m^2q^3\le c d^{1/4} \log^{-2} d$, each extremal maximal-entropy Gibbs state is characterized by an interval of size $m+1$ in which most spins take values. This was previously known only when $m=1$~\cite{peled2010high}.

\smallskip
{\it Notation}.--- For $U\subset\Z^d$, let $N(U)$ be the neighborhood of $U$, $U^+ := U \cup N(U)$, $\partial U$ the set of edges between $U$ and $U^c$, and $\intextB U$ the endpoints of $\partial U$.
$U$ is called odd (even) if $U \cap \intextB U$ consists of odd (even) vertices.
The positive constants $C,c$ may change between lines.

\smallskip
{\it Proofs}.---We describe the proof for the AF Potts model at zero temperature (proper $q$-colorings), as it already contains the essential ideas and is technically simpler to present (full details will appear elsewhere~\cite{peledspinka17}). We take $q\le c\,d^{1/20}\log^{-2/5}d$ so that (GAP) is satisfied. The proof is an involved variant of the Peierls argument, relying on an information-theoretic inequality of Shearer to control entropy loss, and a coarse-graining technique for odd sets to control the entropy of domain walls.

We work in a finite box $\Lambda\subset\Z^d$ containing the origin~$\zero$. Fixing a dominant pattern $P_0=(A_0,B_0)$, we let $f$ be a uniformly sampled proper $q$-coloring of $\Lambda^+$ subject to the constraint that $v$ follows $P_0$ for all $v\in\extB\Lambda$. Our goal is to show that $\zero$ then follows $P_0$ with high probability.

\smallskip
{\it Step 1: breakup}.---In the first step, we identify ordered and disordered regions in the configuration. We aim to associate a subset $X_P\subset\Z^d$ with each dominant pattern $P=(A,B)$ with the idea that the vertices in $X_P$ are ordered according to the $P$ pattern. Our definitions allow the $X_P$ to overlap. A first naive idea is to say that $v\in X_P$ if $v$ follows the $P$ pattern. However, every vertex follows several different dominant patterns and this will not lead to a useful notion of ordering. It turns out to be more useful to \emph{associate} a vertex with the $P$ pattern if the neighbors of the vertex follow the $P$ pattern (though the definition of $X_P$ is still more complicated).

\begin{figure}
		\includegraphics[scale=0.4]{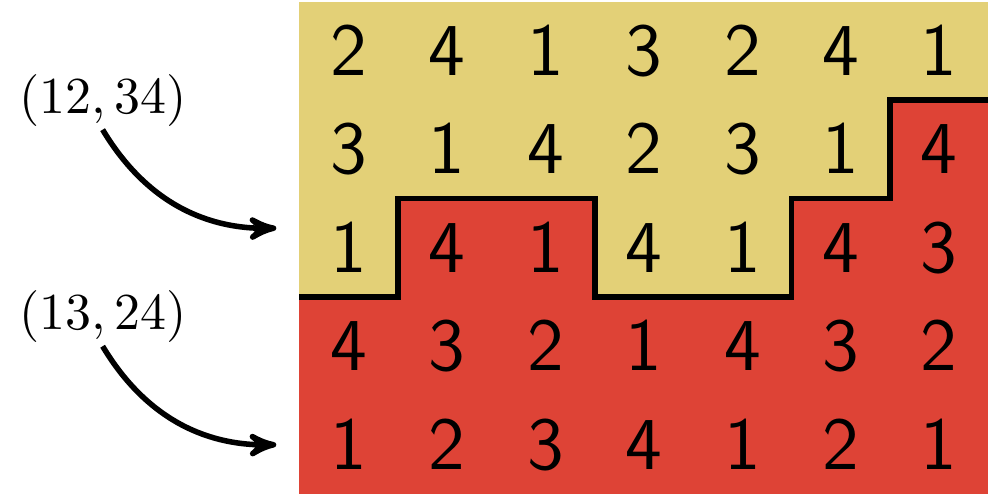}\quad
		\includegraphics[scale=0.4]{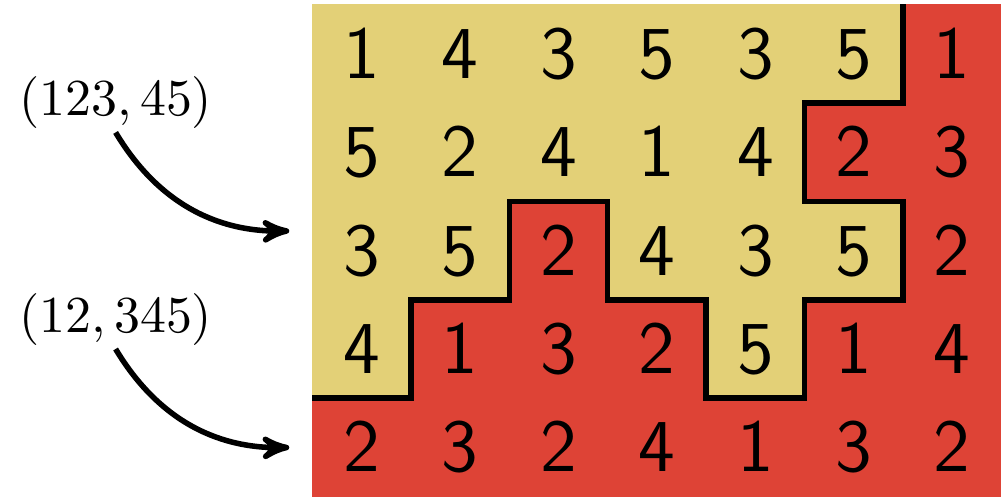}
	\caption{An interface between two regions associated to different dominant patterns (left: $q=4$, right: $q=5$).}
	\label{fig:interfaces}
\end{figure}

Forming an interface between regions $U$ and $U^c$ associated to different dominant patterns leads to a loss of entropy. One checks that the entropic loss is at least $|\intextB U|\log(\frac{q-2}{q})$ for even $q$ and at least $\frac{1}{2d}|\partial U|\log(\frac{q-1}{q+1})$ for odd $q$. In the odd $q$ case, this bound is attained only when $U$ is even or odd, with the smaller side of its dominant pattern on its boundary (see Fig.~\ref{fig:interfaces}). As $\frac{1}{d}|\partial U|\le |\intextB U|$, the odd $q$ case is more delicate. With this in mind, we shall define the $X_P$ to be even or odd (for any $q$), and this choice will be of use also in step~3 below.

Let $P=(A,B)$ be a dominant pattern with $|A|\le |B|$ ($|A|>|B|$). Let $I_P$ be the set of odd (even) vertices associated to $P$ and set $Y_P := I_P^+$. The resulting $Y_P$ is necessarily even (odd). The entropy of the configuration is restricted in several regions: the boundaries $\intextB Y_P$, the regions of overlap $Y_\overlap := \bigcup_{P \neq Q} (Y_P \cap Y_Q)$ and the outside regions $Y_\bad := \bigcap_P (Y_P)^c$. Together, these yield
\begin{equation}\label{eq:Y_star}
  Y_* := (\cup_P \intextB Y_P) \cup Y_\bad \cup Y_\overlap .
\end{equation}
A domain wall for the AF Potts model is a connected component of $Y_*$ \footnote{For technical reasons, connectivity to distance $10$ is used.}. Domain walls may encompass a significant volume due to $Y_\bad$ and $Y_\overlap$, and their complement may consist of several connected components.
The definitions ensure that the vertices in each connected component of $Y_*^c$ are associated to a single dominant pattern.

As we are only interested in the domain walls surrounding $\zero$, we set $X_*$ to be the outermost connected component of $Y_*$ surrounding $\zero$. The \emph{breakup} is the collection $X=(X_P)_P$ defined by letting $v\in X_P$ if $v\in X_*\cap Y_P$ or if $v\in X_*^c$ and the boundary of the connected component of $v$ in $X_*^c$ is ordered according to $P$. We also set $X_\bad:=X_*\cap Y_\bad$ and $X_\overlap:=X_*\cap Y_\overlap$, so that \eqref{eq:Y_star} remains true with $X$ in place of $Y$.

\smallskip
{\it Step 2: entropic cost of the domain wall}.---Say that $X$ has \emph{size} $(L,M)$ if $|\cup_P \partial X_P|= L$ and $|X_\bad \cup X_\overlap|=M$. Our next goal is to show that any specific breakup $\mathfrak{X}$ of size $(L,M)$ is entropically unfavorable:
\begin{equation}\label{eq:given_breakup_bound}
  \Pr(X=\mathfrak{X}) \le \exp\left(- \tfrac{c}{q^2} \big( \tfrac{L}{d}+ \tfrac{M}{qd^{1/4}} \big)\right) .
\end{equation}
To establish this, we apply the following one-to-many operation to every configuration having $X=\mathfrak{X}$: (i) Delete the spin values at all vertices of $X_*$. (ii) For each connected component $D$ of $X_P\setminus X_*$, apply the bijection $\varphi$ taking $P$ to $P_0$ (given by (SYM)) to the spin values at $D$, and also, if $\varphi$ maps $A$ to $B_0$, shift the configuration in $D$ by a single lattice site in the $(1,0,\ldots,0)$ direction \cite{dobrushin1968problem}. (iii) Fill spin values following $P_0$ in all remaining vertices.

Noting that the resulting configuration is always a proper coloring, and that no entropy is lost in step (ii), it remains to show that the entropy gain in step (iii) is much larger than the entropy loss in step (i). The gain in step (iii) is either $\log\lfloor\frac{q}{2}\rfloor$ or $\log\lceil\frac{q}{2}\rceil$ per vertex according to its parity. We proceed to bound the loss in step (i).

We make use of a basic information-theoretic inequality, first used in a similar context by Kahn~\cite{kahn2001entropy}, followed by Galvin--Tetali~\cite{galvin2004weighted}. Let $H$ denote Shannon entropy.

\noindent{\it Shearer's inequality~\cite{chung1986some}:}
Let $Z_1,\dots,Z_n$ be discrete random variables. Let $\cI$ be a collection of subsets of $\{1,\dots,n\}$ such that $|\{I \in \cI : j \in I\}| \ge k$ for every~$j$.
Then
\[ H(Z_1,\dots,Z_n) \le \frac{1}{k} \sum_{I \in \cI} H((Z_i)_{i\in I}) .\]

Let $\Omega$ be the set of configurations having $X=\mathfrak{X}$ and let $f\in\Omega$ be uniformly chosen. Let $F$ be the configuration coinciding with $f$ on $X_*$ and equaling a new symbol $\star$ on $\Z^d \setminus X_*$. Let $\cE$ and $\cO$ be the even and odd vertices in $\Z^d$. Applying Shearer's inequality to $(F_v)_{v \in \cE}$ with $\cI = \{ N(v) \}_{v \in \cO}$, yields
\begin{align*}
H(f_{X_*}) &= H(F)
 = H(F_{\cE}) + H(F_{\cO} \mid F_{\cE})\\
 &\le \sum_{v \in \cO} \left[ \tfrac{H(F_{N(v)})}{2d}  + H\big(F(v) \mid F_{N(v)}\big) \right].
\end{align*}
Averaging this with the inequality obtained by reversing the roles of $\cO$ and $\cE$ shows that $H(f_{X_*})$ is at most
\begin{equation*}
   \frac{1}{2}\sum_{v} \bigg[ \underbrace{\tfrac{H\big(F^{N(v)}\big)}{2d}}_{I} + \underbrace{\tfrac{H\big(F_{N(v)}\mid F^{N(v)}\big)}{2d} + H\big(F(v) \mid F^{N(v)}\big)}_{II} \bigg],
\end{equation*}
where $F_U, F^U$ denote the values of $F$ on $U$ taken as an ordered vector or as an unordered set, respectively. Of course, the terms corresponding to vertices $v$ at distance $2$ or more from $X_*$ equal zero as $F$ is deterministic in their neighborhood. The advantage of this bound is that it is local, with each term involving only the values of $F$ on a vertex and its neighbors. Each term admits the simple bound $I\le \frac{q\log 2}{2d}$ and $II\le\log(\lfloor\frac{q}{2}\rfloor\lceil\frac{q}{2}\rceil)$, which only takes into account the fact that $F(v)\notin F^{N(v)}$. The main contribution to $I+II$ comes from the possibility that for some dominant phase $(A,B)$, $F^{N(v)}=B$ and $F(v)$ and $F_{N(v)}$ are approximately uniformly distributed in $A$ and $B^{2d}$, respectively. To improve upon this, we use additional information implied by having $X=\mathfrak{X}$, captured by the following notions.

A vertex $v$ has \emph{unbalanced neighborhood} in $f$, if some value $i \in f^{N(v)}$ appears in $f_{N(v)}$ at most $d^{3/4}$ times \footnote{The number $3/4$ is picked after some optimization. Other powers less than one lead to similar bounds.}. A directed edge $(v,u)$ is \emph{restricted} in $f$ if $(S_{v,v},S_{v,u})$ is not a dominant pattern, where
\begin{align*}
	S_{v,w} := \big\{g(w) : g \in \Omega,~ g^{N(v)} = f^{N(v)} \big\},
\end{align*}
i.e., if knowing $f^{N(v)}$ guarantees that $v$ and $u$ cannot take all possible values of a dominant pattern.
A vertex $v$ has a \emph{unique pattern} if there exists $A \subset S$ such that for every $g \in \Omega$, either $g^{N(v)}= A$, or $v$ has unbalanced neighborhood in $g$, or all edges $(v,u)$ are restricted in $g$.

Letting $p_v$ denote the probability that $v$ has unbalanced neighborhood in $f$ and $r_v$ denote the expected number of restricted edges $(v,u)$ in $f$, we obtain the improved bound,
\begin{equation*}
  I + II \le \frac{Cq}{d} + \log\left(\left\lfloor\frac{q}{2}\right\rfloor\left\lceil\frac{q}{2}\right\rceil\right) - \frac{c}{q^2} \left(p_v + \frac{r_v}{d}\right).
\end{equation*}
A more modest improvement is obtained by knowing that $v$ has a unique pattern, leading to
\begin{equation*}
  I+II\le e^{-cd/q^2} + \log\left(\left\lfloor\frac{q}{2}\right\rfloor\left\lceil\frac{q}{2}\right\rceil\right).
\end{equation*}
It is easily checked that every vertex in $X_* \setminus X_\bad$ has a unique pattern. Therefore, the entropy of $f_{X_*}$ will be sufficiently small to imply the bound \eqref{eq:given_breakup_bound} when
\begin{equation}\label{eq:sum_of_probabilities}
\sum_{v \in X_*} (p_v + \frac{r_v}{d}) \ge \frac{cL}{d}+\frac{cM}{qd^{1/4}}.
\end{equation}
Every edge between two different $X_P$ and every edge incident to $X_\overlap$ is necessarily restricted in $f$. Unfortunately, $X_\bad$ need not contain enough vertices having unbalanced neighborhood or many restricted edges - the main reason being that $X_\bad$ may contain isolated even (odd) vertices associated to a dominant pattern $(A,B)$ having $|A|\le|B|$ ($|A|>|B|$).
However, there exists a subset $V(f)\subseteq X_\bad$ of size $|V(f)|\le CMd^{-3/4} \log d$ such that revealing $V(f)$ and $(f^{N(v)})_{v \in V(f)}$ increases the number of restricted edges in $X_\bad$ enough to make \eqref{eq:sum_of_probabilities} hold. The entropy of this additional information is negligible with our assumptions.

The above discussion does not take into account the entropic contribution to $I + II$ of vertices at distance $1$ from $X_*$. These can be handled with careful bookkeeping at the boundary vertices $\intextB X_*$. We do not elaborate on this point further as it does not rely on new ideas.

\smallskip
{\it Step 3: structure of breakups and coarse-graining}.---It is temping to conclude that domain walls are unlikely by summing the bound \eqref{eq:given_breakup_bound} over all possible breakups. This approach applies to spin systems in which the ratios $\rho_p, \rho_i$ and $\frac{1}{\rho_a}$ appearing in (GAP) are sufficiently close to $0$, as this leads to an improvement in the analogous bound to~\eqref{eq:given_breakup_bound}. Unfortunately, it does not apply to the AF Potts model, as the number of possible breakups of size $(L,M)$ exceeds the reciprocal of the bound \eqref{eq:given_breakup_bound}. Here, an analysis of the structure of possible breakups is required, as we discuss now.

Consider first the case in which a single droplet of the dominant pattern $P$ is inside a sea of $P_0$, i.e., $X_{P'}=\emptyset$ for $P'\notin\{P_0,P\}$, $X_\bad =X_\overlap=\emptyset$ and $X_P$ is connected, with connected complement. The crucial feature of $X_P$ is that its boundary vertices all have the same parity. The number of such droplets with $|\partial X_P|=L$ boundary plaquettes grows as $2^{(\frac{1+\varepsilon_d}{2d})L}$ for $L$ large \cite{feldheim2016growth}, with $2^{-2d}\le\varepsilon_d\le\frac{C\log^{3/2} d}{\sqrt{d}}$. This is to be contrasted with a similar count when the boundary parity constraint is removed, i.e., the number of connected sets with connected complement and $L$ boundary plaquettes, which grows faster, as $e^{\frac{c\log d}{d} L}$~\cite{lebowitz1998improved,balister2007counting}. The different growth rates are indicative of a deeper structural difference. A typical connected set in high dimensions with no parity constraints scales to integrated super-Brownian excursion \cite{lubensky1979statistics, slade1999lattice}, while a typical odd set has a well-defined macroscopic shape (e.g., an axis-parallel box) on which it adds microscopic fluctuations. This is akin to the breathing transition undergone by random surfaces~\cite[Section~7.3]{fernandez2013random}. This phenomenon has been exploited in previous works~\cite{galvin2004phase,peled2010high,feldheim2015long} to provide a natural coarse-graining scheme for odd sets, grouping them according to their rough macroscopic shape, which have significantly less entropy in high dimensions (of order $L \big(\tfrac{\log d}{d}\big)^{3/2}$) than the odd sets themselves. We proceed in the same manner here, extending the previous schemes from droplets to breakups. The remaining task is then to bound the probability of the breakup belonging to a given group and this is achieved with a suitable modification of step~2. This leads to the Peierls-type estimate,
\begin{equation*}
\Pr\left(\text{$X$ has size $(L,M)$}\right) \le  \exp\left(- \tfrac{c}{q^5 \log d} \big( \tfrac{L}{d}+ \tfrac{M}{qd^{1/4}} \big)\right) .
\end{equation*}
Standard techniques, pioneered by \cite{gallavotti1972equilibrium}, allow to pass from such estimates to the characterization of all equilibrium states. Further information is obtained such as the exponential decay of truncated correlation in each of the extreme equilibrium states.

\smallskip
{\it Acknowledgements}.---We thank Roman Koteck\'y, Ron Lifshitz, Moshe Schwartz, Robert Shrock and Alan Sokal for useful discussions during the preparation of this letter. Research of both authors was supported by the Israel Science Foundation
grant 861/15 and the European Research Council starting grant 678520 (LocalOrder). Research
of Y.S. was additionally supported by the Adams Fellowship Program of the Israel Academy of
Sciences and Humanities.

\bibliographystyle{amsplain}
%\nocite{*}
\bibliography{biblio}

\end{document}